# Considering Time in Designing Large-Scale Systems for Scientific Computing

**Nan-Chen Chen[1], Sarah S. Poon[2], Lavanya Ramakrishnan[2], Cecilia R. Aragon[1,2]**
[1] Department of Human Centered Design & Engineering, University of Washington, Seattle, USA
[2] Data Science and Technology, Lawrence Berkeley National Laboratory, Berkeley, USA
nanchen@uw.edu, SSPoon@lbl.gov, LRamakrishnan@lbl.gov, aragon@uw.edu

**ABSTRACT**
High performance computing (HPC) has driven collaborative science discovery for decades. Exascale computing platforms, currently in the design stage, will be deployed around 2022. The next generation of supercomputers is expected to utilize radically different computational paradigms, necessitating fundamental changes in how the community of scientific users will make the most efficient use of these powerful machines. However, there have been few studies of how scientists work with exascale or close-to-exascale HPC systems. Time as a metaphor is so pervasive in the discussions and valuation of computing within the HPC community that it is worthy of close study. We utilize time as a lens to conduct an ethnographic study of scientists interacting with HPC systems. We build upon recent CSCW work to consider temporal rhythms and collective time within the HPC sociotechnical ecosystem and provide considerations for future system design.

**ACM Classification Keywords**
H.5.m. Information interfaces and presentation (e.g., HCI): Miscellaneous.

**INTRODUCTION**
Computation and data have become cornerstones of scientific discovery [17]. The social and organizational aspects of building the large-scale, distributed resources have long been one of the major interests in the field of computer-supported cooperative work (CSCW). A critical component of the sociotechnical ecosystem of scientific discovery is high performance computing (HPC) or supercomputing. A supercomputer has a very high-level computational capacity and thousands of nodes with powerful networks that connect the nodes together allowing for communication across the nodes.

The next generation of supercomputers, the so-called exascale computing platforms, is currently being designed and the first machines are expected to be deployed in 2022. There are many open questions as to how these machines will be utilized by teams of scientists in the coming years.

There has been little research that specifically looks at exascale or close-to-exascale HPC systems and how scientists work with them today. As these large-scale machines will support significant scientific work and collaboration at scale, it is important to examine the challenges people are encountering that limit the effective use of these machines. HPC centers have a significant impact on scientific discoveries. The National Energy Research Scientific Computing Center (NERSC) is one such supercomputing center. In the last five years, NERSC has produced an average of 1,500 journal publications per year based on computations performed at the center. On average, NERSC users publish ten journal cover stories per year in publications such as Nature. Four Nobel prizes have been awarded for work that used NERSC resources to date [35].

Time as a metaphor is so pervasive in the discussions and valuation of computing within the HPC community that it is worthy of close study. This temporal metaphor is ubiquitous within the community in multiple ways: supercomputers use performance as the primary efficiency metric where performance measures the amount of work done in unit time. Floating point operations have been considered to be central to scientific computing; system performance has historically been measured in 'flops' (floating point operations per second). The design and purchase decisions for these multi-million-dollar systems often revolve around temporal performance metrics such as the LINPACK Benchmark [50].

Research and discussion into the nature of time and its perspectives has a decade-long history in CSCW and other related fields [3, 29-31, 33]. In this paper, we particularly focus on what has been termed *collective time*, specifically as used in collaborative scientific work. [29, 30, 45]. We consider the positioning of time as a unit of analysis that extends beyond the individual and includes both users and the HPC system. In the process, we also build on a large body of CSCW research into temporal rhythms [7, 9, 18, 19, 25, 37, 41-43, 48, 51] for, as Jackson et al. [19] have argued, "distributed collective practices not only have rhythms, but in some fundamental sense are rhythms." Thus we seek to uncover implications for design for the HPC community that take into account temporal rhythms and collective time. Finally, we argue that selective transparency of various rhythms is essential to present an appropriate view of collective time. In this, we follow Goffman [15] and others [29, 30, 40] who discuss how the presentation of a front stage can selectively obscure individual notions of time and focus attention on collective time.

Lindley [29] has noted the dichotomy of human understanding of clock time, contrasting Glennie and Thrift's [14] view of time as "sets of practices, which are bound up with time-reckoning and time-keeping technologies, but which vary and are shaped by different times, places and communities" with the more abstract and mechanistic view of clock time as depicted by Mumford [32] and others. We argue that this dichotomy exists within the HPC community and creates frictions, in the sense that the overt structure of scheduling is currently based on mechanistic and inflexible time while the HPC users fluidly adapt themselves to and repurpose the schedule for their own temporal rhythms, utilizing the system to most efficiently produce the results they need for their work. This can lead to temporal misalignments when an idealized and rigid view of time and efficiency conflicts with the rhythms and end goals of the scientific users.

We utilize time as a lens in our ethnographic study of scientists interacting with HPC systems, a research method proposed by Ancona et al. in 2001 [4] that suggests, by focusing on the temporal aspects, it "makes us speak in a different language, ask different questions, and use a different framework in the methodological aspects of our research. (p.17)" A specific focus on the potential value of considering time as collective rather than individual leads us to identify several conflicting rhythms that arise between individual HPC users and the more monochromic [21] or linear and tightly scheduled view of time imposed upon the community by management: mismatches in time expectations, temporal uncertainties, and conflicting views of optimization.

We unpack the distinctions between mechanistic and rigid rhythms of compute time and consider how time becomes mutable as it is processed, lived with, stretched, and utilized by scientists and software developers who must interact with HPC machines on a daily basis. We claim that to truly understand how best to develop interfaces for the next generation of HPC machines, we need to deeply understand the mental models of the humans who use them, and that those mental models are inextricably bound up with human concepts of time and human concepts of machine time.

The contributions of this work are as follows: via an ethnographic study in a major supercomputing center including participant observation and interviews, we consider and reflect upon the multiplicity of views of time (including collective time) and temporal rhythms in the HPC community, and we further utilize time as a lens to investigate the sociotechnical system of scientific HPC computing. We conclude with a discussion of an interface design that may be able to help HPC users visualize and manage their time collectively based both on previous research in CSCW and CHI [7] and on work in the HPC community [5] on the use of visualization to facilitate situational awareness within technical groups. This leads to potential implications for design of exascale systems that these interwoven and conflicting temporal structures and perceptions have revealed.

## RELATED WORK

Time, or temporality—the experience of time and the temporal organization of activities [43]—has been one of the essential parts of our everyday experiences and working environments. Indeed, large computational systems such as supercomputers were invented to speed up computational work in order to save very critical and precious time. This also explains why, historically, performance has been used as the key evaluation metric. However, as time is a broad topic, for the purposes of this paper, we wish to delve more deeply into the concept of *collective time* for the HPC ecosystem, and to do so, we must consider three strands of CSCW and CHI research: temporal rhythms, collective time, and sociotechnical studies of scientific collaborations.

### Temporal Rhythms in CSCW and HCI

Temporal rhythms in collaborative work have been a fruitful area of study in CSCW for many years [9, 18, 25, 37, 41-43, 51]. Orlikowski and Yates [38] suggested that "people in organizations experience time through the shared temporal structures they enact recurrently in their everyday practices" (p. 686). Landgren [27] documented the rhythms present when Swedish fire and rescue workers operate under time pressure; Nilsson and Hertzum [37] studied collective rhythms in home care in Denmark, and numerous studies have analyzed temporal rhythms in health care [6, 9, 12, 41-43]. Bardram drew from his fieldwork in medical care and developed the concept of *temporal coordination*, based on Activity Theory, that emphasizes the temporal aspect of people coordinating their work [6]. Also in the medical work context, Reddy and Dourish found three emergent temporal features—*trajectories*, *rhythms*, and *horizons*—which affect healthcare providers [43]. In particular, their definition of temporal rhythms highlights the recurring patterns of work and how people leverage the characteristic of reoccurrence to deal with events or activities.

All these examples demonstrate that people react to and utilize temporal rhythms in complex ways. People doing collaborative work have different experiences with such rhythms, and the design of systems impacts the ways users interact with the system. Several studies on temporal artifacts, such as calendars, have shown how people's behaviors are influenced by those artifacts [39]. There are related discussions on how to support the representation of temporal and social structures in online environments for everyday collaboration [13], or how to better design time representation in systems for time-critical medical teamwork [25]. We now turn to the body of previous work specifically studying temporal rhythms in scientific collaboration.

*Temporal Rhythms in Scientific Collaboration*
Managing and analyzing temporal aspects of scientific collaboration has long been a fertile area of CSCW research. Karasti, Baker, and Millerand's work focused on the short-term and long-term temporal scales in the collaborative development of information infrastructure for scientific collaboration [20]. They identified two temporal orientations, *project time* and *infrastructure time*, and they suggested paying more attention to studying long-term development of such collaborations. Jackson et al. extended this concept and proposed the idea of *collaborative rhythm*, describing the temporal dissonance and alignment in collaborative work [19]. They identified four types of rhythms in joint scientific work: *organizational, infrastructural, biographical,* and *phenomenal*. *Organizational* rhythms include those set by institutions, such as academic calendars and funding deadlines. *Infrastructural* refers to the temporal constraints set by the equipment and infrastructure of a scientific venture. *Biographical* rhythms are determined by human life circumstances such as family and illness. Finally, *phenomenal* rhythms emerge from the objects or phenomena under study, such as phases of the moon or the seasonal mating patterns of animals.

More recently, Steinhardt and Jackson considered plans [48] and anticipation work [47]. They suggested that planning is an essential part of collaborative scientific work and that plans align rhythms in local working practice [48]. They further suggested that the pathways of project development are guided by *anticipation work*, which are the practices that cultivate and channel expectations of the future [47]. These studies focus more on the relationships between people in the scientific work context, while our work focuses on the entire ecosystem of users and systems, and seeks to draw links between temporal rhythms and the collective temporal experience of the group including their temporal interactions with their computing systems.

**Collective Time**
Mazmanian and Erickson [30] and Lindley [29] have recently studied and called for further examination of collective time. Mazmanian et al. [31] cite the importance of politics and power within temporality from Sharma [45], who takes a political view of time as "tethered and collective."

Lindley positions time as "collective and entangled," noting that the nature of time is inherently multiple and fraught with contradictions, while Mazmanian and Erickson put forth a definition and challenge: they suggest a focus on "the *collective*, not the individual, level of practice within an organization" [30] to manage human availability.

We suggest that the imperative to position time as collective is an outgrowth of many decades of research into the nature and conflicts of temporality. There are multiple threads of research into temporal conflicts and dichotomies [16, 21, 22, 31]. Many researchers, both within CSCW and outside, have observed fundamental dichotomies in human understandings of time. We note that most of these have focused on the individual and individual perceptions, rather than taking a unit of analysis beyond the individual.

**Sociotechnical Studies of Cyberinfrastructure**
The historical background of our work includes sociotechnical studies of cyberinfrastructure, which has been of major interest in CSCW for decades (e.g., [8, 23, 28, 46]). Starting from systems that support scientific work and distributed collaboration, researchers have been studying various social and organizational aspects around cyberinfrastructure. Ribes and Lee in a special issue of the *Journal of Computer Supported Cooperative Work* in 2010 [44] focused on the social aspects of the development of cyberinfrastructure. Our utilization of a time lens on an under-studied sociotechnical ecosystem, the HPC community, enables us to focus on blending several threads of research as discussed previously. In the next section, we provide necessary background information on this unique ecosystem.

**BACKGROUND AND METHODS**
A HPC system consists of a large group of connected computers that function as a coherent system. Along with the people surrounding the machines, a sociotechnical system with unique characteristics is formed. In this section, we describe our research site, data collection, and analysis methods. We also introduce the roles in this ecosystem and the key characteristics of an HPC system. Finally, we provide an exemplar workflow to describe the interaction of various entities in the ecosystem.

**Research Site**
Our research site is a government-funded research center in the USA where scientists use shared HPC machines run by the National Energy Research Scientific Computing Center (NERSC). NERSC is supported by the Department of Energy (DOE) and operates multiple HPC systems. These systems are shared between hundreds of projects. According to the official website [34], more than 5,000 scientists use NERSC to perform scientific research in various disciplines such as material sciences, earth sciences, cosmology, and climate science.

**Data Collection**
Our fieldwork was conducted under an ongoing research project to design next-generation scientific computational environments. The field study focuses primarily on a group of scientists from a single domain. These domain scientists belong to a large project in which they conduct research via simulation and modeling. Other than the core group, we also interviewed domain scientists from four other collaborations that use the HPC system to validate common problems across domains.

During six months of fieldwork, we conducted 26 semi-structured formal interviews (average length one hour each) with 15 people involved with the HPC system. We also engaged in other qualitative ethnographic methods such as direct observation and shadowing over a period of several weeks. The interviewees included four domain scientists, seven computer engineers who support domain scientists, and four members of the HPC facility staff. 13 interviewees were male and two female, representative of the gender distribution of the members of the HPC community. Interviewees were selected via snowball sampling. Their experience with HPC systems ranged from 5 to 25 years.

The data collection occurred in two stages. In the first stage, lasting three months, we entered into the field to familiarize ourselves with the environment and scientists' thoughts and issues about working with HPC systems. We talked to domain scientists in material science, climate science, physics, and cosmology. In this stage, we asked them to describe their scientific work and experiences with HPC systems. Example interview questions included:

- Can you tell me about the projects you are working on that involve the use of supercomputing?
- What is the science objective of the project?
- What kinds of data and computational work are involved?

During this period, time-related comments became more and more salient, so we narrowed down our investigative focus to time-related topics. Some example questions in this stage were:

- When is the last time you felt you spent too much time on one task? What and why?
- How much compute time allocation do you have from NERSC?
- How do you decide what types of jobs to spend your compute time on?
- How do you track the amount of your allocation?

In stage two, we interviewed five domain scientists, five computer engineers, and four HPC facility staff members. Three of the five domain scientists and one computer engineer interacted with us frequently outside of interviews during the six-month period. We had at least four interviews with each of the members of the core group.

**Data Analysis**
We conducted a thematic analysis [10] to analyze the dataset. The first author reviewed each transcript, and time-related quotations were pulled out. Subsequently, we coded the quotes in three passes. We first open-coded the issues around the quotes. Then, we organized the quotes into different stages of a job cycle and synthesized the codes into themes in each stage. Finally, we generated higher-level themes from the codes for discussion.

**Roles in the HPC Ecosystem**
In our field study, the people involved in using computational cycles on HPC systems fall into the following key roles[1]: domain scientists, computer engineers, and HPC facility staff.

*Domain Scientists*
Domain scientists are researchers in specific areas of basic science, such as cosmology, microbial biology, material science, and climate science. Most domain scientists conduct their work as part of one or more scientific collaborations. One or more principal investigators (PI) lead the collaboration. Team members often include senior scientists, mid-career scientists, and early-career postdocs and students. In order to answer fundamental science questions, they run codes (software) written either by their scientific community or by their collaborators. The scale of the codes and/or the data associated with them require these to be run in computing environments larger than the typical workstation, such as in cloud, cluster, and HPC environments. Domain scientists are the primary users of the HPC system we studied.

*Computer Engineers*
Computer engineers are members of the scientific collaboration who help scientists run codes on the HPC system. This work may involve modifying existing software or 'codes' to utilize the systems, installing community codes, writing custom codes to aid in the running and analysis of scientific codes, and developing cyberinfrastructure to support the scientific workflows.

*HPC Facility Staff*
HPC staff is comprised of people in various roles surrounding the acquisition and support of HPC systems. This includes people who act as liaisons between the users of the system and the facility to understand users' needs, people who manage the allocation of system resources among its users, and people who work directly with users to help them troubleshoot their system usage.

**Key Characteristics of an HPC System**
Since HPC systems comprise a group of interconnected computers, using an HPC system is different from using a workstation. Here, we briefly describe a subset of characteristics related to using these systems.

*Programming for HPC*
Each of the individual computers in an HPC system, called nodes, contains several processors (or cores) and some amount of memory that is shared among the processors. A serial application is one that executes calculations in a specific sequence by a single processor. In order to take advantage of the thousands of processors in these systems, often applications are parallelized—calculations are executed simultaneously on different processors, allowing more calculations to be processed in a given timeframe.

---
[1] This is not an exhaustive list of roles involved in the entire network but only the key roles on which we have focused in this paper.

These supercomputers usually have a variety of storage options including large scratch space, which is local to the system, and global storage for user and project directories. Additionally, users have access to tape storage for long-term archival needs. The process of reading and writing files to these storage systems by an application is generally referred to as I/O. Applications can generate large amounts of data, sometimes larger than the amount of disk space a user can store locally. It may be necessary to move data off local storage to larger disk and tape storage systems.

Parallelization and data management require additional coding and understanding of how to utilize the system. Scientists often consult and work with computer engineers and facility staff to learn how to write and configure software for these tasks.

### Allocations

In order to use a system such as NERSC, a scientific collaboration must submit a proposal to DOE to request a specific amount of computational time for the coming year. The PI on the proposal will usually be the main point of contact. Since collaborations often share a single allocation, the PI will work with the other scientists to determine the appropriate amount of time to request. The time requested is in units of compute hours, which is typically an hour of time for a single processor. Thus, determining the amount of allocation to request requires estimating how much time a collaboration will need to conduct specific science goals. The collaboration will work with computer engineers and facility staff to make this estimation.

Based on several factors, including the projected impact of the intended science, the DOE and the facility managers determine the size of the allocation of compute hours to the proposing collaboration, which may be less than the amount requested.

### Batch Queues

Once a scientific collaboration has its allocation, scientists then need to actually schedule time for their application to run on the supercomputer. The scientist must submit the application as a job into a queue, requesting the number of processors to be reserved and the maximum amount of time requested on the system. The time that a job runs on a system is called *wall time*, and an application that exceeds the requested wall time will be terminated. Based on the order of job submissions and the priority of the jobs, an automated scheduler will optimize the utilization of the system. The goal of the scheduler is to maximize usage, i.e., the scheduler tries to allocate processors as much as possible throughout the day. Most jobs submitted to the queue will not be able to run immediately on submission. Typically, the queue wait time increases as more processors and longer wall times are requested. The number of processors reserved and the number of hours used to run the application are two of the main factors that determine how many compute hours of the total allocation are used by each job. Another important factor that affects both queue wait time and allocation charge is the type of queue. For example, jobs in the priority queue can take priority over jobs in the standard queue. The facility will charge more against the allocations to run jobs in the priority queue. There are several types of queues at a facility, and the facility staff will often set policies around these queues based on discussions with various facility users. The science users, with help from computer engineers and facility staff, will then decide which queues to use based on their needs and how their applications run.

**Exemplar Workflow in the HPC Ecosystem**

Several of the scientific collaboration teams we studied are composed of both domain scientists and computer engineers. Running scientific codes at an HPC facility will involve people in all the roles listed earlier. Here we describe an exemplar workflow to illustrate the interplay of these roles in the HPC ecosystem.

Collaborations use models that encapsulate the physics of natural systems. This process is called simulation. The output data of these simulations are not in the format needed to analyze the data, so the collaboration creates programs to convert it into the desired format (post-processing). Finally, the data are analyzed using custom-built analysis tools. Throughout the process, data are moved through different storage systems. Published data include data consumed by others in the community (publishing) and data meant for long-term preservation (archiving).

Getting this workflow to a functional state on an HPC system is called preparation. For example, codes may need to be modified so they can be executed in a batch queue system—required libraries may need to be installed on the machines and codes may need to be compiled and parallelized. Computer engineers work with the domain scientists to accomplish these tasks. User support staff at the facility provide aid for some of these tasks as well and also provide additional guidance on how to use the various components of an HPC system.

**System Design Trade-Offs**

HPC system design is a complex and often unwieldy process that includes multiple stakeholders from different organizations including HPC facility staff, funding agencies, hardware and software vendors. System design of these large-scale systems is influenced by various factors including projected user needs as extrapolated from day-to-day user support and user requirements workshops, hardware options available through the vendors, and cost. However, the vendors who provide the hardware and software often have limited exposure to the users. Additionally, performance and utilization are often the primary metrics of success in these environments; past design decisions have not focused on usability or the user perspective. With the move to exascale computing, however, there has been acknowledgment of the need to take usability into account due to the potential magnitude of inefficiencies in system usage.

## FINDINGS: TEMPORAL RHYTHMS AND CONFLICTS IN THE HPC SOCIOTECHNICAL SYSTEM

Time is at the center of interactions within the HPC sociotechnical system. In this section, we report the results from our study. We consider the various temporal characteristics, rhythms, and conflicts across the lifecycle of a scientist's use of an HPC system. We structure the following section via areas of scientific use of HPC resources, and identify three cross-cutting temporal themes: mismatches in time expectations, temporal uncertainties, and conflicting views of optimization. In this and the discussion section, we build on Jackson et al.'s research on collaborative rhythms to call out commonalities and conflicts.

### Time Allocation

Access to supercomputers is shared across different user groups from various science communities, and a project is allocated compute time and data storage on a yearly basis. Time becomes a virtual currency in this environment, with the amount allocated to a project being shared between project members. Time is deducted from the allocation on an ongoing basis. Additionally, different queues and systems have different "costs" and priority access to machines costs extra as well [36]. The number of nodes and cores a HPC system has is fixed, putting an upper bound on the "resources" available in this economic system. It is important to note that having virtual currency does not guarantee immediate access to the resource. Even if a project has a large allocation, project members may still have to wait in a queue for other jobs to finish on the machines.

An important question is how do users decide on the amount of "time" they request? Generally, this is guided by the scientific goal the scientist wants to achieve. However, in addition to the obvious "machine time" that will be needed to run the models and analyses, users might consider other "human time issues." For example, scientists might consider their own time and how much they can finish in the proposed timeline of the project.

*In terms of deciding how many simulations I am going to run in a given year, my decisions over the past couple of years have not been governed by how long it takes things to run on the supercomputer, but rather how much time it takes me to manage these runs versus giving me time to do analysis.* [Domain Scientist B]

These contradictory views of human and machine time optimization represent conflicts between what Jackson et al. [19] term biographical and infrastructural rhythms. Additionally, the distribution of allocation creates interesting dynamics in the economic system. For example, some scientists reported that the allocation was less than what they asked for and is not sufficient to accomplish their scientific goals. On the other end of the spectrum, some projects used less of their allocation than originally requested.

*We asked for 62 million [hours], but only got 50 million hours in the end. We should have asked for 75 million hours because we really need 62 million hours.* [Domain Scientist A]

This mismatch in time expectations creates stress and a perception of inefficiency and unfairness, which staff work hard to address. The virtual aspect of currency allows for redistributions. Supercomputer staff assess usage and conduct adjustments quarterly to ensure machines are heavily utilized. Projects that use less than their predicted amounts might lose some of their allocation. Partial amounts of unused allocation will then be redistributed to other projects. Therefore, some scientists experienced an increase in their time allocation to compensate for their heavy use as they used more resources:

*I had requested 15 million hours and I had been granted 12 million hours, and at the end I had 15 million ... the quota was incrementally increased just ahead of us each time.* [Domain Scientist B]

Here, a mismatch in time expectations has been resolved due to manual intervention. Both initial distribution and redistribution of "machine time" involves informal interactions between the HPC center staff and the domain scientists, and includes a complex manual decision process. The HPC center staff juggle complex temporal rhythms, attempting to meet scientists' needs while maintaining high system utilization.

### Preparation Stage

*Human Time to Get Codes Running on the HPC System*
Scientists have to invest time to port their work to supercomputers since HPC machines are substantially different from traditional workstation environments. Domain scientists invest time in learning the system and are usually able to perform this task by themselves using online tutorials. This time investment is usually expected to pay dividends, in terms of being able to run codes faster and/or scale to problem sizes that would not be possible without access to these supercomputers.

The interactions in the sociotechnical system become more interesting as the complexity of the codes and/or scale increases. Scientists need a deep understanding of system hardware and software to obtain best utilization of HPC resources. This involves a steeper learning curve and hence a larger time investment. Scientists optimize the use of their own time in various ways: some learn parallel programming in depth while others gain only a surface understanding. Computer engineers and/or HPC center staff spend time helping scientists prepare and optimize their codes at various levels. Time spent improving codes can result in better performance (i.e., less machine time used).

*Data Management Time*

HPC environments provide various storage options, including temporary scratch space, project file systems, and archival tape storage. The file system space on the HPC system is limited, so they have to store their data on tapes and transfer them back to the file system when they want to conduct analysis on the data. Fetching the data they need from tape potentially takes a long time. Sometimes, the size of the data may exceed the amount of project space they have, so they store their data in the scratch/temporary directory since it has more space. Periodically, the HPC facility staff members sweep away all files in the temporary directory, which has important time implications for the domain scientists. Scientists need to move the data they want to save off the temporary space and subsequently bring back any data they need for future processing.

Computer Engineer: *It looks like they are going to clean up the temporary directory next week. Please backup your important files.*

Domain Scientist A: *I wonder how long it takes to read 15 terabytes of data from tape.*

Computer Engineer: *Probably a long time.*

Domain Scientist A: *Probably a long time. I bet it takes a week.*

This conflict between organizational and infrastructural rhythms [19] led to annoyance and the perception of wasted human time. Thus, we see that domain scientists often associate "time" estimates to the work items they do with respect to the HPC system.

**Queue Waiting Time**

Once a domain scientist has prepared the codes to run on the HPC machine, they wrap their codes in a script and submit a "job" to the HPC system.

*Queue Waiting Time for Code Development*

Scientists understand they have to wait for their jobs to run on the machine. However, scientists definitely expressed a level of annoyance at having to wait in the queue for a quick test during code development or testing. When scientists are still developing their codes there may be bugs that cause the program to crash. In order to debug the code, they will have to go through a long wait in the queue multiple times. In HPC systems, there is a special queue called the debug queue — i.e., short jobs with lesser waiting time — which is reserved for testing.

*The turnaround when you are running something through the queue system, even in the debug queue, can be quite a while. You don't have the rapid turnaround that you do at a work station or something like that, where you test something, change it, test something, and change it.* [Domain Scientist C]

These contradictory views of optimization represent conflicts between infrastructural and biographical rhythms: even short jobs in the debug queue have to wait for what is considered a long time by the scientists.

*Uncertain Queue Waiting Time*

The domain scientists understand they have to wait to use the system, but they are troubled by the uncertainty of the waiting time and unpredictability of when their jobs might finish. The only feedback the system gives them is the position in the queue. However, this does not tell the scientists the time the jobs ahead of them in the queue might take. The queue number also does not always indicate the order of execution due to varying priorities and policies associated with the jobs.

*I'll submit a job and it goes in and it is at number 1300 and it starts an hour later. Then in another case I will have a job which might be there for a day or half a day and it is down to 100 in the queue apparently, and the next time I look it's at 1500, then it starts.* [Domain Scientist B]

This temporal uncertainty created by misalignment between institutional and biographical rhythms leads to frustration and the perception of inefficiency. Since it is hard to tell when a job might start, scientists report either spending many hours babysitting the jobs actively or forgetting the jobs that were previously submitted.

The HPC queue system provides a "showstart" command to enable users to estimate the waiting time. The command displays the *earliest possible start time* of jobs that request the same amount of resources. However, this is not a personalized estimate. The documentation clearly notes that jobs requesting the same amount of resources will return the same start time. Thus, only the highest priority job with the same characteristics will start at the give time. Just like the queue position indicator, it does not provide accurate information to tell you how long a job has to wait.

*I won't call it an estimate since estimate is a too strong of a word. I really do not find it to be successful. It does give you an idea if a job is going to take a long time. But it might take 20 minutes or an hour.* [Domain Scientist C]

*Scheduling Policies*

As described in the background section, every job to the HPC machine is submitted to a queue to wait for its turn for execution. The scheduler then decides the execution order. It considers a variety of factors, such as the current load of the machines, the number of processors the job requests, the priority of the project, the number of jobs a user has in the queue at the moment, and the maximum time a job is allowed to run, (called "wall time"). Typically, jobs are scheduled in a first come first serve order. However, when there are holes in the schedule, a lower priority job might fill the gap. The priority of a job is determined by the project it belongs to and queue it resides in (e.g. regular queue vs. high-priority queue.) Like all scheduling problems in computer science or other domains, scheduling on the HPC system is an NP-hard problem, which means it is difficult to compute an optimal solution to the problem.

Thus, heuristics are used to find a solution. The policies around scheduling are complex, and even the HPC facility staff acknowledge that it is hard to predict the scheduling outcome.

*The scheduler's policies are very complicated. It is very difficult to figure out when your job is going to run. That's one thing people always want to know. That is one thing we can never really tell them.* [HPC Facility Staff Member A]

Conflict between infrastructural and biographical rhythms is perceived by and acknowledged as frustrating by staff members as well.

The complexity of the scheduling is further increased by other mechanisms in the system that allow for higher priority needs. Users can "spend" twice the allocation to get a higher priority on the machine. Sometimes, a job may be about to start when another user submits a job to the high priority queue. Thus, the new job starts before the original one that was about to run.

HPC facility staff accept special requests to boost priority of the jobs or even block out a period of time for a certain project. One of our interviewees, a domain scientist, reported to us that he got extra help from the HPC facility staff to speed up his waiting time in the queue:

*For the biggest job, the queuing time takes almost a week because I'm running on about sixty thousand processors. The facility staff was kind enough to give me a temporary boost for a couple months, but I don't think they will give me that again for a while.* [Domain Scientist C]

### Queue Time vs. the Ways of Working
At first glance, the cost in human time of code running could be pegged to the amount of wait time and clock time, which might be seen as pure inefficiency. But is the cost as high as it appears? Perhaps during this wait time, the scientist has time to reflect upon a previous run, generate the type of creative insight that often occurs during downtime, or accomplish other critical tasks. Thus there may be unexpected hidden value in this "waiting" time.

*While I'm waiting for one, I can be setting up another...if the queues were faster, then you wouldn't work that way. You would actually maybe just work serially instead of trying to multi-task yourself, which is an efficient way to use the time, but not necessarily your head.* [Domain Scientist C]

This raises some important issues that need to be considered in the design of systems. Scientists adapt their own temporal rhythms (and work methods) to work around temporal conflicts and inefficiencies in the systems. As is well known in the field of CSCW [1, 24, 28, 29], this adaptation may yield unexpected dividends, or may have hidden costs.

## Execution Time
A run might involve various steps from environment initialization to actual process execution. In this section, we discuss temporal factors related to the execution phase.

### Time Cost in Initialization Stage
Traditional HPC applications used Fortran/C/C++ for their codes and HPC systems are optimized for these languages. Increasingly, we see the use of Python for scientific computing. In our studies, the scientists mentioned problems with using Python for jobs that required a large number of nodes. Python imports all the dependent libraries on each node, and each importing step involves a file system read. When 4,000 nodes are trying to access the same set of files, it impacts performance. A number of participants reported that it took at least an hour each time just to initialize the environment, before any analysis or simulation started running. Thus jobs not only take longer to run, but the users get charged for this initialization time that is often unaccounted for in allocation requests.

Problems such as these and the resulting frustrations often drive scientists to find ways to improve the situation. We observed one scientist who was impacted strongly by this problem develop workarounds with support from computer engineers. The solution minimized I/O overhead. The HPC center staff is also investigating a more permanent solution to the Python problem in collaboration with the vendors. It is interesting to note that the temporal rhythms of these interactions are different. A scientist might be able to come up with a workaround for a problem specific to their application in a short time, as in hours or days. However, a system-wide solution requires systematic design, implementation and testing. This can take months, sometimes even years. This conflict between infrastructural and phenomenal rhythms [19] raises some important questions for the HPC community to consider in the context of collaborative work: Would this have been possible without the timely collaboration between the computer engineers and the scientists? Is there a way to create an ecosystem that allows users to share ideas and solutions in a way that harmonizes these differing temporal rhythms? Can we establish a synergy between user-specific workarounds and long-term solutions?

### Variability in Running Time
The HPC batch queue system requires users to set "wall time" as an estimate of the actual running time in terms of clock hours. If a job runs longer than the preset wall time, it will be killed by the system. However, the actual running time is highly dependent on the performance of the underlying systems and possible load on shared resources. One typical case is I/O: since filesystems are shared between nodes, one highly demanding job using the file system may slow others down considerably.

*When I try and run something on the HPC system, sometimes just even opening a text file can take a really long time, depending on if someone is overutilizing the*

*system or really banging on I/O on the system that I'm trying to run on. So I guess that is a consideration I have had to keep in mind, or learn to keep in mind in moving stuff from a desktop to the HPC system. I/O can be really variable.* [Domain Scientist C]

This temporal uncertainty impacts scheduling. Users often pad their jobs and specify higher wall clock times due to the uncertainty in running time.

*For the low resolution simulations, they typically finish in about 20 minutes, but because of the volatility of NERSC I have had them take 45 minutes so I specify an hour. I think that all of them sit for an hour except for the high resolution ones which take substantially longer. Those I give three hours. Because I have seen them take between an hour and a half and two and a half or something. In order to avoid failure completely, I am going quite a bit higher than I think it needs. Because at this point, I don't have a good system for dealing with job failure.* [Domain Scientist C]

This is a common phenomenon across users of the NERSC systems. A previous analysis of the workload showed that 60% of the jobs only need half of the wall time they requested [2]. A higher wall clock time might mean a longer wait time in the queue, since shorter jobs might be able to backfill. However, if a job runs over the specified wall time, it gets killed and results in lost work and lost time (work time, waiting time and allocation time).

*Wasted Time from System Uncertainties*
Large-scale systems such as supercomputers often experience various types of failures. A single failure on one node can impact the entire application that might be running on 1000 nodes. It is often hard to tell what actually happened in one of the computational nodes among 10,000 nodes. Therefore, it is challenging for a scientist to identify whether the problem is in the code or the system. Additionally, transient errors might rectify themselves and the cause of the problem might be unknown. Our participants mentioned that there were times when a job failed the first time and then succeeded the second time.

*You don't always get the same result when you do something twice… Sometimes I will run something literally without changing anything, resubmit the same job again. It will have failed once. It will run successfully the second time.* [Domain Scientist C]

These types of errors can lead to interesting user behavior. Scientists might repeatedly try to run the same job to identify if it was a transient error, effectively wasting machine time. Alternatively, a user might invest time to debug and try to understand the problems, which might be a wasted effort.

**Time to Optimize**
Scientists often conduct various kinds of optimizations, from algorithm optimization to performance optimization. How does a scientist determine where to invest their time and energy? As one of the scientists pointed out, it is often not a case of obtaining the best performance but getting sufficiently good performance:

*I am not really interested in making a script that takes an hour, run in 10 minutes. I am interested in taking a script that runs three days, and running in one, or less ... Where my interests are, is making the intractable problem, tractable; not making the tractable problems faster, because they're tractable, who cares?* [Domain Scientist A]

In another example, the scientist didn't think it was worth his time to learn something new to parallelize the code. Human time investment is often based on what causes them frustration and/or what is important from a science perspective, and not necessarily from the system perspective, as has long been noted in CSCW research [1].

*This loop could be parallelized. I don't know how to do that inside a loop. I know how to parallelize over files. I don't know how to parallelize over loops with the OpenMP calls. I could learn it but I haven't.* [Domain Scientist A]

However, significant waste of system time will often cause users to consider optimizations. In one particular case, a scientist reported that he had trouble with configuration parameters that caused a significant slow-down. The slow-down in turn had the possibility of draining their allocation a fair bit. Thus the scientist spent time trying to figure out the cause of the performance slow-down, and eventually had to obtain help from collaborators in another institution to fix the problem.

*I could have used the time and just burned a lot of CPU time, but I was getting a factor of 100 slower performance. It seemed to me a waste of CPU time to do that.* [Domain Scientist C]

**Time to Handle System Upgrades**
During the period we conducted the fieldwork, the operating system on the HPC system was upgraded. The upgrade broke many software dependencies of our participants' existing codes.

*Every time there's an operating system upgrade, it hurts us badly. We haven't gone through any of them without some kind of scar. Sometimes it's really bad. This one is really bad. It may be weeks or months before we actually can run again.* [Domain Scientist A]

Even though this strongly impacts the users, this time cost is also hard to prevent. Usually before the upgrades, the HPC facility staff will test the base packages extensively. However, it is hard to guarantee all the software packages will work because the users often manage some of their domain specific tools which might have complex dependencies.

*More issues do happen after upgrades, because the systems and the software are so complicated that it is not uncommon. There are so many interactions it is not*

*uncommon for some interactions to have not been fully tested by somebody somewhere if something happens.* [HPC Facility Staff Member A]

Sometimes, scientists might choose to introduce steps in the processing that add time to their process in order to deal with the obstacles they have. For instance, after the OS upgrade broke the model, the scientists introduced another step in the computation workflow, thus increasing the machine time.

*To deal with the problem [the OS upgrade], I had to move my process to another file system so I was not getting this error. Of course copying the data over and everything takes a bit of time, and it is also just introducing a new layer, a new step.* [Domain Scientist B]

Humans make their decisions based on perceived time. Will it cost them more in terms of personal or learning time to switch to a new model than it does to recompile? People make their computer usage decisions by framing these questions within personal time cost and biographical rhythms.

**Collaboration Time**
Domain scientists are theoretically able to save time by asking computer engineers for help with programming and running codes. However, we found from the interviews that scientists may not necessarily use the help.

One scientist noted that it would take him more time to explain it to the computer engineers. This is a good example of the scientist's perspective of cost versus benefit and a classic CSCW trade-off. While it would take the scientist some time and effort to do the work, the process of seeking help would be more cumbersome.

*It is something where I felt I couldn't ask the computer engineers to do it because of the amount of time it would take me to explain the details to him. I could just as easily have written that myself.* [Domain Scientist B]

**DISCUSSION**

**The Nature of Sharing**
Many aspects of time and temporal rhythms in our fieldwork are tied to the nature of shared resources, thus confirming our belief that a unit of temporal analysis beyond the individual makes sense in this domain. For example, because the system is shared between users and the demands from users are more than what the system can supply at a time, jobs have to sit in queue to wait for others. Resources such as storage are also shared and, hence, have quotas associated with them. Sharing is integral to the system due to the costs associated with it. To shape a group's perception of resources as a common good, past research has demonstrated that understanding time as collective could lead to a more effective mental model of both human and machine time [5, 7, 13].

**Temporal Rhythms**
A recurring theme we identified was the conflict between various temporal rhythms, as has been noted by Jackson et al. [19] within other scientific collaborations. For example, in order to make the code run faster, the scientists may have to spend time on improving it through the use of new libraries or techniques. When human time is required to save machine time, this conflict between biographical and infrastructural rhythms requires trade-offs. Improved run time may lead to benefits for the scientists in the long term. But these benefits may not be obvious or what the scientists are interested in as we saw earlier. If the problem is already tractable, scientists have little interest in speeding it up.

A more complex example is when a scientist spends time in order to optimize the configurations of the job for queuing time and running time. This effort can be tedious due to the complexity of the codes and delays in the queuing system. Therefore, scientists tend not to invest time on it unless the issues become a source of difficulty for them. In some cases, they may turn to find external help from computer engineers or HPC center staff. This introduces an interaction between scientists' time and computer engineers' time to save machine time. According to the interviews, sometimes this interaction is critical and helpful, as computer engineers may be able to solve scientists' issues in a short time. However, sometimes scientists feel that the overhead of explaining the details to someone else is not worth it. Then, they prefer to solve the problems by themselves or just do the tasks manually. This classic CSCW problem has been addressed by multiple researchers over decades. The trade-off between individual vs. collective needs has been addressed at length in the field (e.g. [1, 24, 49]). Based on previous work, we suggest that a consideration of collective time may prove useful in addressing this complex and nuanced problem.

**Challenges in Communication**
We found that some issues in time arise when members of the ecosystem face challenges in communicating their state or intentions. The system does not always provide useful accounts or explanations, of its activities. For example, scientists felt that the queue order appeared uncorrelated to the queue wait time. The opaqueness of the system caused the scientists to spend time babysitting the job. If the queue wait time could more predictably be determined, scientists may take other actions instead, such as deciding to go home or wait for the results. Yet, as the facility staff explained to us, even if the system was completely transparent, accurate queue wait times are difficult to determine, due to factors such as the policies around queues and how users use the system.

Challenges in communication were observed not just between the system and people but also between people. For example, as seen earlier, the scientists described the time overheads in back and forth conversation with the computer engineer explaining the details of their work and scripts. All these examples show that communicating state and intentions is challenging and complex, as previous work has demonstrated [1, 5-7, 11, 25, 27, 30, 31]. This body of research has also shown that increased visibility of state and intentions, especially if focused collectively, is likely to be beneficial.

**Misaligned Objectives**

Different members of the HPC ecosystem have different objectives. This misalignment can lead to conflicts in temporal rhythms. We have found that facility staff and computer engineers often have a strong interest in the most efficient use of the system. However, the main objective of scientists is to get their work done, and this does not always result in efficient system use. For example, we saw earlier that users often pad their wall times. Facility staff strongly encourage their users to provide accurate wall times, as this results in better scheduling of jobs and presumably better user experience. However, for some users, crossing the wall time and having their job terminated early can lead to significant time and effort trying to restart their job. Even when they generally know how long their application will run, unpredictable system state can lead to much variability in the run time, leading to even larger wall time padding.

Given how Lindley and other CSCW researchers [1, 29, 38] have noted that technology can give shape to the ways in which time is organized, some of the frictions and conflicting rhythms surfaced by our study may be addressed through technological affordances, for example, visualizations of the collective rhythms of HPC user and compute time, that provide a selective transparency into collective scheduling of time.

**Developing a Scheduling Interface for Collective Time**

Historically, system performance is the most commonly used design guideline for HPC systems. Yet, when considering the ecosystem as a whole, there exist opportunities for developing new types of metrics for HPC systems. Our ethnographic time lens revealed that system performance is only one of several ways that time affects both the machines and the humans in the HPC ecosystem. We suggest the possibility of developing a set of metrics based on collective time. In this, we build on the foundation laid by Lindley [29] and Mazmanian and Erickson [30] in their qualitative discussions of collective time. For example, optimizing code for HPC may save machine time in terms of run time but comes at a cost to domain scientists in terms of education and implementation time. Other time costs may come in the form of communication with computer engineers and facility staff. This example illustrates how time shapes the experiences within the ecosystem and uncovers the possibility of considering collective time as a design guideline for HPC interfaces.

We follow Jackson [19] in that designing for collective time requires designing for temporal rhythms. Similarly, as Kuutti and Bannon remark in their view on the practice turn in HCI, "Practices are wholes, whose existence is dependent on the temporal interconnection of all these elements, and cannot be reduced to, or explained by, any one single element." Further, "the individual user cannot be the unit of analysis… Practices are a shared resource among a community of people." [26]

To address the issue of collective time in the HPC community, we turn to previous CSCW and CHI research. In 2002, Begole et al. [7] described a system that provides visualizations of patterns of users' computer activity data in the form of a set of stacked activity plots which are aligned vertically in a series of horizontal bar graphs.

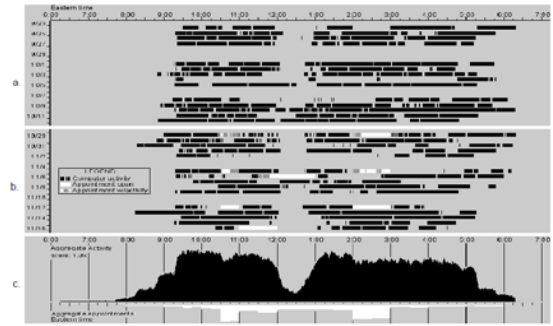

**Figure 1. Begole's visualization of human activity on computers in a collaborative group [7].**

This technique was successfully used for group coordination, as a way to maintain awareness of other people's activity patterns and work rhythms. We suggest a similar collective visualization of both human activity and the HPC schedule/allocation data, color-coded by user group, might serve to highlight these work rhythms and enable collective efficiency. Creating such a visualization of HPC allocations along with its users could enable more efficient use of the system, as Begole showed their visualization was helpful in predicting availability and inactivity for human work rhythms and it may apply also to the sociotechnical applications of collective time that include a shared resource and human time constraints. Selective disclosure and transparency of information would be necessary (due to privacy concerns). Begole suggests that visualizing work rhythms could lead to a means of creating a shared sense of time within a workgroup.

Other lines of research indicate this approach could be effective. Aragon et al. [5] found that collaborative visualizations contributed to a shared sense of awareness and alignment of temporal rhythms within a scientific group operating under time pressure. Fisher et al. [13] noted that making temporal structure visible led to increased coordination within a work group, and that visualizing temporal patterns was especially helpful. They warned about the potential risk of loss of privacy, which is why we emphasize selective transparency of individual rhythms and a focus on visualizing collective time, in other words utilizing metrics based on a unit of analysis beyond the individual to create such a visualization. Further work is needed to refine and validate such metrics, and this proposed design obviously needs a significant amount of study before it can be realized. Nevertheless, we suggest that multiple lines of research are converging in this area and thus we echo Lindley's [29] and Mazmanian and Erickson's [30] call for more research into collective time. We further suggest that HPC and exascale environments may be excellent venues for studies of such designs, as the stakes are very high, as are the potential rewards. Given the tremendous impact of centers such as NERSC on worldwide scientific discovery, the impending risks and benefits of the move to exascale computing, and the potential of CSCW research to deliver useful knowledge to collaborative HPC ecosystems, these converging lines of research hold significant potential.

## CONCLUSION

Via a qualitative study of users of HPC systems and the application of a time lens to identify issues surrounding these large, sociotechnical ecosystems, we built on previous research on temporal rhythms in collaborative scientific work and the emerging concept of collective time to suggest design implications for HPC and potential directions for future work. We found that time and temporal rhythms play an important role at every stage of computational work. Time is not simply passively experienced, but rather actively shapes the dynamics in this sociotechnical system. We discussed how the essence of a shared system imposes both support and limitations on resources, leading to a complicated balancing act between supply and demand. We examined the trade-offs between various types of temporal rhythms in the context of Jackson et al.'s [19] work, and identified three sources of temporal conflict: mismatches in time expectations, temporal uncertainties, and conflicting views of optimization.

We illustrated how the misaligned objectives and rhythms and the challenges of communication between people and machines and people with each other might be resolved through consideration of collective time. We have identified some areas to be considered when systems are designed. We suggest that effort be devoted to foregrounding collective time, not just traditional performance benchmarks, and to developing visualizations that allow selective transparency of human and machine time within a collective framework. HPC has enabled many scientific discoveries, but understanding how its community of users interacts with its mechanisms from a sociotechnical and temporal perspective will be essential for the next generation of scientific breakthroughs.

## ACKNOWLEDGMENTS
This work was funded by the Office of Science, Office of Advanced Scientific Computing Research (ASCR) of the U.S. Department of Energy under Contract Number DE-AC02-05CH11231 and award number DE-SC0012474.